\documentstyle[preprint,epsfig,aps]{revtex} 
\def\ra{\rightarrow}
\def\be{\begin{equation}}
\def\ee{\end{equation}}
\def\bea{\begin{eqnarray}}
\def\eea{\end{eqnarray}}
\def\lln{\left<}
\def\rln{\right>}
\begin{document}
\preprint{BARI-TH/432-2002}
\title{REMARKS ON CP ASYMMETRIES IN $D^0/\overline {D^0} \ra K_S \pi^+ \pi^-$}

\author{S. Arunagiri\footnote{electronic address: shen.arunagiri@ba.infn.it}}

\address{Dipartimento di Fisica - Gruppo Teorico \\
Universit\'a di Bari, Via Amendola 173, Bari, 70126, Italy}

\date{\today}

\maketitle

\begin{abstract}
We consider the interference of resonant amplitudes leading to 
the final state $K_S \pi^+ \pi^-$ in $D^0/\overline {D^0}$ decays. Each of 
these amplitudes
consists of both Cabibbo allowed and doubly Cabibbo suppressed transitions.
The role of strong phase arising out of Breit-Wigner resonant propagators is 
emphasised.
Invoking the $\Delta S = \Delta Q$ rule, $K_S$ in the final state is identified
as the mass eigenstate of superposed weak eigenstates $K^0$ and $\bar K^0$.
A nonzero CP asymmetry appears to be possible in three body decays.
\end{abstract}

\newpage

As immense activities are on at charm/B factories searching for CP violating 
effects, the study of weak dynamics of charm sector acquires renewed interest. 
The CP asymmetry occurs if there exist nonvanishing
weak and strong phases provided by a pair of amplitudes which are
distinct by both phases with respect to one another, through interference
\cite{sandip}. For instance, in the two body $D^0$ decays into $K^\pm \pi^\mp$, 
the strong phase arises due to strong interaction effects such as rescattering 
and final state interaction. But, such a strong phase is small, for example 
it is about $13^\circ$ in the model of Buccella {\it et al} 
\cite{buch}, thus resulting in a CP asymmetry of $O(10^{-3})$\cite{bigi0,nelson}.  

Alternatively, one can look at three body modes. The three body
final state is reached by more than one way,
namely, nonresonant and resonant modes. There are many resonant
modes:
That is, $D^0 \ra M_1R \ra M_1[M_2M_3]_R$, where $R$ stands for 
resonance and $M$'s for mesons. Given two resonant modes that lead to
the same final state, there arise distinct strong phase between them 
due to the Breit-Wigner (BW) resonant propagator and the angular momentum 
quantum numbers of the resonance and of its decay products \cite{soni}.
Thus, a strong phase is nontrivial. 

Besides, the correlation between the net strangeness (produced)\footnote{In 
the strange decays, this is change in strangeness.} and the net charge 
(involved), namely, the $\Delta S = \Delta Q$ rule. Unlike in strange decays 
where violation strangeness-charge symmetry means the mixing of $K_S$ and 
$K_L$ \cite{sanda}, in charm decays into strange final state, this 
rule qualifies $K_S$ is the mixed state of the weak eigen states $K^0$ and
$\bar K^0$. This rule implies the coherent superposition of CA and DCS 
transitions.
 
In this letter, we look at the significance of strong phase that arises
due to BW propagators of intermediate resonances which lead 
to a common final state in $D^0$ three body decays,
as an extension of the Atwood and Soni's proposal for $B$ decays\cite{soni} 
to charm sector and the importance of invoking the 
$\Delta S = \Delta Q$ rule thereof.
For $D^0 \ra K_S \pi^+ \pi^-$, we consider the resonant modes of 
$K^*(890)^\pm, \rho^0(770)$ and $f_0(990)$:
\bea
D^0 &\ra& K^{*-} \pi^+ \ra \bar K^0 \pi^+ \pi^-\\
D^0 &\ra& \bar K^0 \rho^0 \ra \bar K^0 \pi^+ \pi^-\\
D^0 &\ra& \bar K^0 f_0 \ra \bar K^0 \pi^+ \pi^-
\eea
These resonances show up in the Dalitz plot
 of BABAR data \cite{palano}.

Let us write the amplitude for the resonant decay $D^0 \ra K_S \pi^+
\pi^-$ mediated by a resonance $i$ as 
\be
M_i = A_i \Pi_i B_i
\label{ampl1}
\ee
where $A_i$ is the weak part of the amplitude containing the CKM phase,
$\Pi_i = [s-m_i^2+i \Gamma_i m_i]^{-1}$ the BW propagator that provides
the strong phase and the
strong coupling $B_i = (16 \pi m_i^3 \Gamma_i
/\lambda^{1/2}(m_i^2,m_1^2,m_2^2))^{1/2}$. 
In order to extract the strong phase, we rewrite $\Pi_i$ as
\be
\Pi_i = \tilde{\Pi}_i e^{i \delta_i}
\ee
The strong phase $\delta$ is given by the width and mass of the resonance.

We take into account four amplitudes due to them: 
both the CA and DCS transitions of each.
The amplitude for $D^0 \ra K_S \pi^+ \pi^-$ is then expressed after factoring 
out the strong and weak phases as
\bea
M &=& e^{i \gamma_1} \left[ e^{i \delta_i} M_i^C + e^{i \delta_j} M_j^C \right] +
e^{i \gamma_2} \left[ e^{i \delta_i} M_i^S + e^{i \delta_j} M_j^S \right] \label{ampt}\\   
\overline {M} &=& e^{-i \gamma_1} \left[ e^{i \delta_i} 
\overline {M}_i^C + e^{i \delta_j} 
\overline {M}_j^C \right] + 
e^{-i \gamma_2} \left[ e^{i \delta_i} \overline {M}_i^S + e^{i \delta_j} 
\overline {M}_j^S \right]    
\eea 
where $\gamma_{1,2}$ stands for the CA and DCS weak phase, $\delta_{i,j}$
the strong phase and the superscript $C$ and $S$ for CA and DCS transitions.
The amplitude $|M|$ is resultant of coherent superposition of the four amplitudes\cite{bigi}. 
As notation, $K^*$ is identified with $i$ and $\rho^0$
and $f_0$ with $j$.   

Then we have the asymmetry as
\be
a_{CP} = {2W\sin \Delta \sin \phi \over 
{1+X+4Y\cos \Delta +2Z\cos \phi
+2W\cos \Delta \cos \phi}}
\label{acp}
\ee
where $\phi = |\gamma_1 - \gamma_2|$ is the weak phase and 
$\Delta = |\delta_i-\delta_j|$ the strong phase and
\be
W = R_3+R_1R_2, \hspace{0.2in}
X = R_1^2+R_2^2+R_3^2, \hspace{0.2in}
Y = R_1+R_2R_3,\hspace{0.2in}
Z = R_2+R_1R_3
\ee
with $R_1, R_2$ and $R_3$ respectively the ratio of $M_j^C, M_i^S$ and $M_j^S$
with respect to $M_i^C$:
\bea
R_1 &=& {f_\rho \lln \bar K^0 |(\bar sc)_{V-A}|D^0\rln 
\over {f_\pi\lln K^*|(\bar sc)_{V-A}|D^0\rln }} {\hat {\Pi}_\rho B_\rho \over
{\hat{\Pi}_{K^*}B_{K^*}}},\hspace{0.3in}
R_2 = {f_{K^*} \lln  \pi |(\bar dc)_{V-A}|D^0\rln 
\over {f_\pi\lln K^*|(\bar sc)_{V-A}|D^0\rln }}, \nonumber\\
R_3 &=& {f_{K^0} \lln  \rho^0 |(\bar dc)_{V-A}|D^0\rln 
\over {f_\pi\lln K^*|(\bar sc)_{V-A}|D^0\rln }} {\hat {\Pi}_\rho B_\rho \over
{\hat{\Pi}_{K^*}B_{K^*}}}
\label{rs}
\eea
where in (\ref{rs}) factorisation approximation is applied for the weak amplitude.
Similarly, $R$'s are for $f_0$. We note caution that not much is known learly
about $f_0$. However, we have treated this on par with $\rho^0$. The width
of these resonances are 50, 150 and 50-100 MeV respectively of
$K^*(890)$, $\rho^0(770)$ and $f_0(980)$. In this note, for calculational
purpose, the width of $f_0$ is chosen as 75 MeV.

\begin{figure}
\begin{center}
 \epsfig{figure=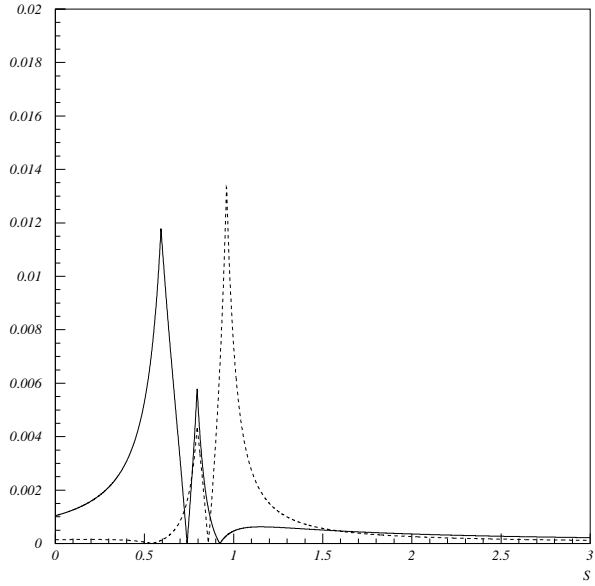,width=3.5in}
 \caption{CP asymmetry Vs. $s$: $\rho^0$ and $K^*$
  (solid line), $f_0$ and  $K^*$ (dashed line). 
  The CP asymmetry exhibits minimum, sign change and maximum as 
 a function $s$. The weak phase $\phi = 0.4 \times 10^{-3}$.
 }
 \label{fig:acp}
 \end{center}
\end{figure}

The strong phase $\Delta$ is determined as a function of $s$. As an order of 
magnitude, the weak phase $\phi$ is chosen as $0.4 \times 10^{-3}$ as
$\phi \sim arg(V_{cd}V_{us}^*/V_{cs}V_{ud}^*)$. 
The ratios of the weak matrix elements in (\ref{rs}) are assumed to be $O(1)$
as they may turn out to be so in the $SU(3)$ limit. The CP asymmetry is shown in
Fig. (\ref{fig:acp}) as a function of $s$. The strong phase $\Delta$ is 
about $57^\circ$ for $K^*$ and $\rho^0$ and about $64^\circ$ for $K^*$ and 
$f_0$ at $s = m_{K^*}^2$. At $s = m_{K^*}^2$, the amplitudes are in phase 
with respect to one another, yielding constructive interference of each pair 
leading to the same final state. 
The minima of $a_{CP}$ for $K^*-\rho$ and $K^*-f_0$ occurs respectively
at $s = 0.75$ and $0.85$ GeV$^2$ respectively, whereas the maxima are 
respectively at $m_{\rho}^2$ and $m_{f_0}^2$.

$\Delta S = \Delta Q$ rule:  
At quark level, $D^0 \ra K_S \pi^+ \pi^-$ proceeds via 
$c \ra s \bar d u (q \bar q)$, internal $W$-emission, and 
$c \ra d \bar su (q \bar q)$, external $W$-emission, 
corresponding to the CA and DCS transitions respectively. While
the net charge $\Delta Q$ is zero in both, the strangeness production
$\Delta S$ is $-1$ and $+1$ respectively.  
Thus, for $D^0 \ra K_S \pi^+ \pi^-$, the $\Delta S = \Delta Q$ is satisfied
by (\ref{ampt}) if and only if the mass eigenstate $K_S$ in the final state 
is the superposition of the weak eigenstates $K^0$ and $\bar K^0$. 
In other words, in order to reach this final state, the coherent superposition
of the CA and DCS transitions is the basic requirement. 

In the case of $D^\pm \ra K_S \pi^\pm \pi^0$, the intemediate resonances play
similar role in giving rise to a strong phase. As noted in \cite{lip},
the mixing of final state neutral $K$ mesons brings in an additional phase factor
of $3.3 \times 10^{-3}$. The role of this extra contribution to the CP asymmetry
would eventually be significant.

To conclude, we considered the effects arising out of the BW propagators.
These effects are large and thus lead to a possible 
large CP asymmetry. Application of the strangeness-charge symmetry rule 
reveals the structure of $K_S$ in the final state and is expected to 
shed more light on the possible new physics which would be responsible for a 
reasonable CP violating effects. A precise determination of CP aymmetry
depends on the contribution of angular momenta of the resonces and of their 
decay products.

\acknowledgements
The author is greatly indebted to P. Colangelo and A. Palano for useful 
discussions, S. Pakvasa for fruitful comments and P. Majewski for 
helps. 

\end{document}